\let\oldvec\vec
\documentclass[lnicst]{svmultln}
\let\vec\oldvec
\usepackage{makeidx}  
\usepackage[cmex10]{amsmath}
\usepackage{amsfonts}
\usepackage{cite}

\usepackage[dvips]{graphicx}
\graphicspath{{graphics/}}
\DeclareGraphicsExtensions{.eps}

\usepackage{url}
\usepackage[pdfpagelabels,hypertexnames=false,bookmarksopen=true,bookmarksopenlevel=2]{hyperref}

\usepackage{array}

\usepackage{mdwmath}
\usepackage{mdwtab}

\DeclareMathOperator{\erf}{erf}

\begin{document}
\mainmatter              
\title{Improving Diffusion-Based Molecular Communication with Unanchored
Enzymes}
\titlerunning{Molecular Communication with Unanchored Enzymes}  
%
\author{Adam Noel\thanks{The first author was supported
by the Natural Sciences and Engineering Research Council of Canada, and a
Walter C. Sumner Memorial Fellowship. Computing resources were provided
by WestGrid and Compute/Calcul Canada.}
\and Karen Cheung \and
Robert Schober}
\authorrunning{Adam Noel et al.}   
%
\tocauthor{Adam Noel, Karen Cheung, Robert Schober}
\institute{Department of Electrical and Computer Engineering\\
University of
British Columbia, Vancouver BC, Canada\\
\email{adamn@ece.ubc.ca}}

\newcommand{\dbydt}[1]{\frac{d#1}{dt}}
\newcommand{\pbypx}[2]{\frac{\partial #1}{\partial #2}}
\newcommand{\psbypxs}[2]{\frac{\partial^2 #1}{\partial {#2}^2}}
\newcommand{\dbydtc}[1]{\dbydt{\conc{#1}}}
\newcommand{\thev}{\theta_v}
\newcommand{\thevi}[1]{\theta_{v#1}}
\newcommand{\theh}{\theta_h}
\newcommand{\thehi}[1]{\theta_{h#1}}
\newcommand{\x}{x}
\newcommand{\y}{y}
\newcommand{\z}{z}
\newcommand{\rad}[1]{\vec{r_{#1}}}
\newcommand{\radmag}[1]{|\rad{#1}|}

\newcommand{\kth}[1]{k_{#1}}
\newcommand{\km}{K_M}
\newcommand{\vm}{v_{max}}
\newcommand{\conc}[1]{[#1]}
\newcommand{\conco}[1]{[#1]_0}
\newcommand{\C}{C}
\newcommand{\Cx}[1]{C_{#1}}
\newcommand{\CxFun}[3]{C_{#1}(#2,#3)}
\newcommand{\Cobs}{C_{obs}}
\newcommand{\Nobsavgt}{\overline{{\N}_{obs}}(t)}
\newcommand{\Nobsavg}[1]{\overline{{\N}_{obs}}\left(#1\right)}
\newcommand{\Nobsavgmax}{\overline{{\N}_{obs}^\star}}
\newcommand{\Cgen}{C_A(r, t)}
\newcommand{\radbind}{r_B}

\newcommand{\N}{N_A}
\newcommand{\M}{M}
\newcommand{\smM}{m}
\newcommand{\A}{A}
\newcommand{\X}{S}
\newcommand{\metre}{\textnormal{m}}
\newcommand{\second}{\textnormal{s}}
\newcommand{\molecule}{\textnormal{molecule}}
\newcommand{\Dx}[1]{D_{#1}}
\newcommand{\Nx}[1]{N_{#1}}
\newcommand{\Da}{D_\A}
\newcommand{\En}{E}
\newcommand{\Ne}{N_{\En}}
\newcommand{\De}{D_\En}
\newcommand{\EA}{EA}
\newcommand{\Nint}{N_{\EA}}
\newcommand{\Di}{D_{\EA}}
\newcommand{\Etot}{\En_{Tot}}
\newcommand{\stepl}{r_{rms}}
\newcommand{\AP}{A_P}
\newcommand{\Ri}[1]{R_{#1}}
\newcommand{\ro}{r_0}
\newcommand{\rone}{r_1}
\newcommand{\visc}{\eta}
\newcommand{\bolt}{\kth{B}}
\newcommand{\temp}{T}
\newcommand{\T}{T_B}
\newcommand{\Vobs}{V_{obs}}
\newcommand{\robs}{r_{obs}}
\newcommand{\Ve}{V_{enz}}
\newcommand{\Nobs}{N_{obs}}
\newcommand{\tint}{\delta t}
\newcommand{\tmax}{t_{max}}
\newcommand{\Cobsfrac}{\alpha}

\newcommand{\thresh}{\tau}
\newcommand{\Pobs}{P_{obs}}
\newcommand{\Pobsx}[1]{P_{obs}\left(#1\right)}

\newcommand{\fof}[1]{f\left(#1\right)}
\newcommand{\lam}[1]{W\left(#1\right)}
\newcommand{\EXP}[1]{\exp\left(#1\right)}
\newcommand{\ERF}[1]{\erf\left(#1\right)}
\newcommand{\SIN}[1]{\sin\left(#1\right)}
\newcommand{\SINH}[1]{\sinh\left(#1\right)}
\newcommand{\COS}[1]{\cos\left(#1\right)}
\newcommand{\COSH}[1]{\cosh\left(#1\right)}
\newcommand{\Ix}[2]{I_{#1}\left(#2\right)}
\newcommand{\Jx}[2]{J_{#1}\left(#2\right)}

\newcommand{\B}[1]{B_{#1}}
\newcommand{\w}{w}
\newcommand{\n}{n}

\newcommand{\new}[1]{\textbf{#1}}
\newcommand{\ISI}{ISI}

\maketitle              

\begin{abstract}
In this paper, we propose adding enzymes to the propagation environment
of a diffusive molecular communication system as a strategy for mitigating
intersymbol interference. The enzymes form reaction intermediates with
information molecules and then degrade them so that they have a smaller chance
of interfering with future transmissions. We present the reaction-diffusion
dynamics of this proposed system and derive a lower bound expression for the
expected number of molecules observed at the receiver. We justify a
particle-based simulation framework, and present simulation results that show
both the accuracy of our expression and the potential for enzymes to improve
communication performance.
\keywords {Molecular communication, reaction-diffusion system,\\
intersymbol interference, nanonetwork.}
\end{abstract}

\section{Introduction}
Molecular communication is the use of molecules emitted by a transmitter
into its surrounding environment to carry information to an intended receiver.
This strategy has recently emerged as a popular choice for the design of new
communication networks where devices with nanoscale components need to
communicate with each other, i.e., nanonetworks. Molecular communication is
suitable because its inherent biocompatibility can facilitate implementation
inside of a living organism; many mechanisms in cells, organisms, and
subcellular structures already rely on the transmission of molecules for
communication, as described in \cite[Ch. 16]{RefWorks:588}. It is envisioned,
as in \cite{RefWorks:540, RefWorks:608}, that by using bio-hybrid components
(such as synthesized proteins or genetically-modified cells), we can take
advantage of these mechanisms for a range of applications that can include
health monitoring, targeted drug delivery, and nanotechnology in general.

The design of molecular communication systems should reflect both the limited
capabilities of small individual transceivers and the physical environment in
which they operate. The state-of-the-art has only begun to take advantage of the
unique characteristics of molecular communication systems and their operational
environments. The simplest and arguably most popular molecular communication
scheme proposed has been communication via diffusion. Diffusion is a
naturally-occurring process where free molecules
tend to disperse through a medium
over time. Diffusion requires no added energy and can
be very fast over short distances; bacterial cells, many of which are on the
order of one micron in diameter, can rely on diffusion for all of their internal
transport requirements; see \cite[Ch. 4]{RefWorks:587}. By adopting
diffusion, network designers do not need to worry about the development of the
infrastructure required for active methods such as the molecular motors
described in \cite{RefWorks:485}.

The major drawbacks of using
diffusion are the need for a large number of information molecules to send a
single message, long propagation times over larger distances, and the
intersymbol interference (\ISI) due to molecules taking a long time to diffuse
away. Fortunately, biological systems commonly store large numbers of molecules
for release at specific instances, such as the storage of Calcium ions in
cellular vesicles until they are needed for signalling or secretion, as
described in \cite[Ch. 16]{RefWorks:588}. Thus, delay and \ISI\,
become the performance bottlenecks. Strategies in the literature for mitigating
\ISI\, have been limited to making the transmitter wait sufficiently long for
the presence of previously-emitted molecules to become negligible, as in
\cite{RefWorks:512, RefWorks:546, RefWorks:548}. The primary drawback of
this strategy is a reduced transmission rate.

We propose adding reactive molecules to the propagation environment to
significantly decrease the \ISI\, in a molecular communication
link when a single type of information molecule is used. The reactive molecules
transform the information molecules so that they are no longer recognized by
the receiver.
If using chemical reactants, then they must be provided in stoichiometric excess
relative to the information molecules, otherwise their capacity to transform
those molecules may be limited over time. However, a catalyst lowers the
activation energy for a specific biochemical reaction but does not appear in
the stoichiometric expression of the complete reaction so (unlike a reactant) is
not consumed.

An enzyme is a biomolecule that acts as a catalyst, often by
providing an active site (a groove or pocket) that encourages a particular
molecular conformation; see \cite[Ch. 3]{RefWorks:588}.
Compared to catalysts in general, enzymes can have the advantage of very high
selectivity for their substrates.
Thus, we are specifically interested in enzymes as reactive molecules because
a single enzyme can be recycled to react many times.
Enzymes play a key role in many essential biochemical reactions. For example,
acetylcholinesterase is an enzyme present in the neuromuscular junction that
hydrolyzes diffusing acetylcholine to prevent continued activation in
the post-synaptic membrane because the receptor in the membrane does
not recognize acetate or choline, as described in \cite[Ch.
12]{RefWorks:587}.
Acetylcholine is called the \textit{substrate} for acetylcholinesterase. The
physical environment of the neuromuscular junction is referred to as a
\textit{reaction-diffusion system} because reaction and diffusion can take
place simultaneously. From a purely communications perspective, the enzyme in
this example is reducing the \ISI\, of the substrate.

There are many potential benefits for using enzymes to aid in developing new
molecular communication systems. The reduction in \ISI\, would
enable transmitters to release molecules more often, simultaneously increasing
the data rate and decreasing the probability of erroneous transmission. There
would also be less interference from neighbouring communication links, so
independent sender-receiver pairs could be placed closer together than in an
environment dominated by diffusion alone. These gains can be achieved with no
additional complexity at the sender or receiver, which is a very useful benefit
for the case of individual nanomachines with limited computational
capabilities.
The enzymatic reaction mechanism could also be coupled to a mechanism that
regenerates information molecules once they are degraded so that they are
returned to the sender for future use (as is the case for
acetylcholinesterase). Of course, it is necessary to select an enzyme-substrate
pair that would not otherwise damage the environment where the nanomachines are
in operation.

Most existing work in molecular communications,
including \cite{RefWorks:557} and \cite{RefWorks:625}, have considered
enzymes only at the receiver as part of the reception mechanism. In these cases,
the ability for the enzymes to mitigate \ISI\, is limited. Two works that have
considered information molecules reacting in the propagation environment are
\cite{RefWorks:513, RefWorks:469}. In \cite{RefWorks:513}, the spontaneous
destruction and duplication of information molecules are treated as noise
sources, whereas in \cite{RefWorks:469}, information molecules undergo
exponential decay in an attempt to mitigate \ISI.
Papers that have
considered reaction-diffusion systems with enzymes in the propagation
environment from a biological perspective, such as \cite{RefWorks:581} and
\cite{RefWorks:583} for acetylcholinesterase, have focussed on providing an
accurate simulation model for specific biological processes with a particular
physical layout and not the manipulation of parameters for the design of new
communication systems.

In this paper, we present a model for the analysis of diffusion-based
communication systems with enzymes that are present throughout the entire
propagation environment. We start with the fundamental dynamics of both
diffusion and enzyme kinetics to derive a bound on the expected number of
molecules within the volume of an isolated observer placed some distance from
the transmitter. In this context, we assume that the reader has
a communications background and is not
familiar with reaction-diffusion dynamics. We justify a
particle-based simulation framework to assess the accuracy of our
analytical results, and show that adding enzymes drastically reduces the
``tail'' created by relying on diffusion alone.

The rest of this paper is organized as follows. In Section~\ref{sec_model},
we introduce our model for transmission between a single transmitter and
receiver. This model is based on both reaction and diffusion. In
Section~\ref{sec_perf}, we derive the number of information molecules expected
at the receiver.
We present the simulation framework in Section~\ref{sec_sim} before giving
numerical and simulation results
in Section~\ref{sec_results}.
In Section~\ref{sec_concl}, we present conclusions and discuss the
on-going and future direction of our analysis.

Unless otherwise noted, we use meters ($\metre$) for distance, seconds
($\second$) for time, and molecules per $\metre^3$ for concentrations
(concentrations are typically given in moles per litre, but molecules per
$\metre^3$ makes our analysis easier to follow by limiting the number of
conversions).

\section{System Model}
\label{sec_model}

We consider an unbounded 3-dimensional aqueous environment. There is a sender
fixed at the origin, treated analytically as a point source but as a sphere in
simulation. The receiver is a fixed spherical volume of radius $\robs$ and
size $\Vobs$, centered at the point defined by $\rad{0} = \{\x_0,\y_0,\z_0\}$.
The receiver acts as a passive observer by not disturbing the diffusion of any
molecules in the environment. This is not a strong assumption, since many small
molecules are able to diffuse freely through cells and other objects if the
molecules are non-polar or if there are protein channels specific to the
molecules in the cell's plasma membrane; see \cite[Ch. 12]{RefWorks:588}. The
immobility of both the sender and receiver is generally impractical at the
nanoscale unless they are anchored to larger objects, but here we assume
immobility for ease of analysis.

\begin{figure}[!tb]
\centering
\includegraphics[width=8cm]{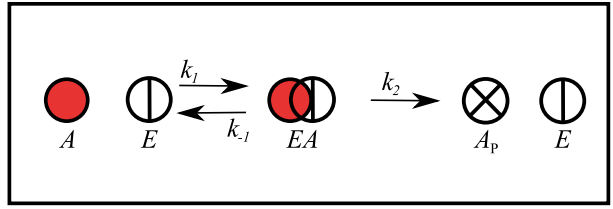}
\caption{The Michaelis-Menten reaction mechanism. Substrate
molecule $\A$ can react with enzyme molecule $\En$ if they collide with
sufficient energy and in the correct orientation. The reaction produces an
intermediate $\EA$ that can either return to its original constituents
or degrade particle $\A$ into $\AP$. The enzyme is not degraded by this
process so it can react with multiple $\A$ molecules. An $\A$ molecule, once
degraded, cannot be returned to its original state via this mechanism.}
\label{enz_rxn_mech}
\end{figure}

Before describing our communication process, we must overview the environment's
chemical dynamics.
There are three mobile species (types of molecules) in the system that we are
interested in: $\A$ molecules, $\En$ molecules, and $\EA$ molecules. The
number of molecules of species $\X$ is given by $\Nx{\X}$ where $\X
\in \{\A,\En,\EA\}$.
$\A$ molecules are the information molecules that are released by the sender.
These molecules have a natural degradation rate that is negligible over the
time scale of interest, but they are able to act as substrates with enzyme
$\En$ molecules.
We assume that $\A$ and $\En$ molecules react according to the following
Michaelis-Menten reaction mechanism (which is generally accepted as the
fundamental mechanism for enzymatic reactions; see \cite{RefWorks:587,
RefWorks:585}, and Fig.~\ref{enz_rxn_mech}):
\begin{align}
\label{k1_mechanism}
\En + \A &\xrightarrow{\kth{1}} \EA, \\
\label{kminus1_mechanism}
\EA &\xrightarrow{\kth{-1}} \En + \A, \\
\label{k2_mechanism}
\EA &\xrightarrow{\kth{2}} \En + \AP,
\end{align}
where $\EA$ is the intermediate formed by the binding of an $\A$ molecule to an
enzyme molecule, $\AP$ is the degraded $\A$ molecule, and $\kth{1}$, $\kth{-1}$,
and $\kth{2}$ are the reaction rates for the reactions as shown with units
$\molecule^{-1}\metre^3\,\second^{-1}$, $\second^{-1}$, and $\second^{-1}$,
respectively. We see that $\A$ molecules are irreversibly degraded by reaction
(\ref{k2_mechanism}) while the enzymes are released intact so that they
can participate in future reactions.
We are not interested in the $\AP$ molecules once they are formed
because they cannot participate in future reactions.

We assume that every molecule of each species $\X$ diffuses independently of all
other molecules, unless they are bound together. We assume that all
free molecules are spherical in shape so that we can state that
each molecule diffuses
with diffusion constant $\Dx{\X}$, found using the Einstein relation as
\cite[Eq. 4.16]{RefWorks:587}
\begin{equation}
\label{JUN12_60}
\Dx{\X} = \frac{\bolt\temp}{6\pi \visc \Ri{\X}},
\end{equation}
where $\bolt$ is the Boltzmann constant ($\bolt = 1.38 \times 10^{-23}$ J/K),
$\temp$ is the temperature in kelvin, $\visc$ is the viscosity of the
medium in which the particle is diffusing ($\visc \approx
10^{-3}\,\textnormal{kg}\,\metre^{-1}\second^{-1}$ for water at room
temperature), and $\Ri{\X}$ is the molecule radius. Thus, the units for
$\Dx{\X}$ are $\metre^2/\second$. The diffusion of a single molecule along one
dimension has variance $2\Dx{\X} t$, where $t$ is the diffusing time \cite[Eq.
4.6]{RefWorks:587}.

We
note that reaction rate constants are experimentally measured for specific
reactions under specific environmental conditions (i.e., temperature, pH, etc.)
using large populations of each reactant.
By ``large'', we mean sufficiently large for the rate of change of species
concentrations to be deterministic. Diffusion also becomes deterministic with
sufficiently large populations. We are not interested in such large populations
due to the size of our system. However, the rate constants also describe the
stochastic affinity of reactions in single-molecule detail, as proven in
\cite{RefWorks:621}. It is impossible to precisely predict where a specific
molecule will diffuse and if or when it will react with other molecules, but the
diffusion and reaction rate constants will be used
to generate random variables when executing stochastic simulations of system
behavior.

\begin{figure}[!tb]
\centering
\includegraphics[width=5cm]{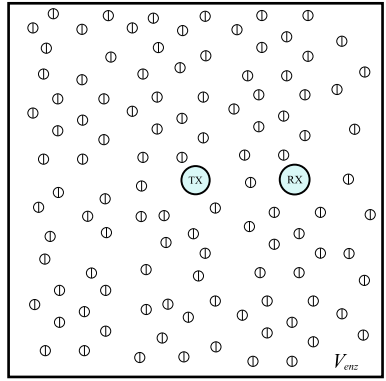}
\caption{The bounded space $\Ve$ in 2-dimensions showing the initial uniform
distribution of enzyme $\En$. $\Ve$ inhibits the passage of $\En$ so that the
total concentration of free and bound $\En$ remains constant. $\A$ molecules can
diffuse beyond $\Ve$.}
\label{venz}
\end{figure}

We can now describe the communication process. The sender emits impulses of $\N$
$\A$ molecules, which is a common emission scheme in the molecular
communication literature; see, for example, \cite{RefWorks:512}.
We deploy binary modulation with constant bit interval $\T$, where $\N$
molecules are released at the start of the interval for binary 1 and no
molecules are released for binary 0 (there have been works studying the use of
different $\T$s depending on the values of the current and previous bit, as in
\cite{RefWorks:548}, since, for example, consecutive 0s can be transmitted with
less risk of intersymbol interference). $\Ne$ $\En$ molecules are randomly
(uniformly) distributed throughout a finite cubic volume $\Ve$ that includes
both the sender (TX) and receiver (RX), as shown in Fig.~\ref{venz}.
$\Ve$ is impermeable to $\En$ molecules (so that we can simulate using a finite
number of $\En$ molecules) but not $\A$ molecules (in simulation, we make $\EA$
molecules decompose to their constituents if they hit the boundary).
Therefore, the total concentration of the free and bound enzyme is constant. $\Ve$
is sufficiently large to assume that it is infinite in size, such that there
would be negligible change in observations at the receiver if $\Ve$ were also
impermeable to $\A$ molecules.

The receiver counts the number of
free (unbound) $\A$ molecules that are within the receiver volume, without
disturbing those molecules. For a practical bio-hybrid system, the $\A$
molecules would need to bind to receptors on either the receiver surface or
within the receiver's volume, but we assume perfect passive counting in order
to focus on the propagation environment. We also assume that the degraded $\AP$
molecules were modified in such a way that they cannot be detected by the
receiver, so $\AP$ molecules can be ignored.

\section{Observations at the Receiver}
\label{sec_perf}

Generally, the spatial-temporal behavior of the three mobile species can be
described using a system of reaction-diffusion partial differential equations.
Even though these equations are deterministic, we noted in
Section~\ref{sec_model} that they will enable stochastic simulation. In
this section, we use the deterministic partial differential equations to derive
the expected number of information molecules at the receiver.

\subsection{Diffusion Only}

For comparison, we first consider the dynamics when there is no
enzyme present, i.e., $\Ne = 0$. So, we only consider the diffusion of $\A$
molecules in the unbounded environment. By Fick's Second Law we have \cite[Ch.
4]{RefWorks:587}
\begin{equation}
\label{APR12_21}
\pbypx{\CxFun{\X}{\rad{}}{t}}{t} = \Dx{\X}\nabla^2\CxFun{\X}{\rad{}}{t},
\end{equation}
where $\CxFun{\X}{\rad{}}{t}$ is the point concentration of species $\X$ at time
$t$ and location $\rad{}$. Closed-form analytical solutions for partial differential
equations are not always possible and depend on the boundary conditions that
are imposed.
Here, we have no $\En$ molecules, so there are also no $\EA$
molecules, and we immediately have
$\CxFun{\En}{\rad{}}{t} = \CxFun{\EA}{\rad{}}{t} = 0 \;\forall\, \rad{},t$.
Assuming that the $\A$ molecules are
released from the origin at $t = 0$, we then have \cite[Eq.
4.28]{RefWorks:587}
\begin{equation}
\label{APR12_22}
\CxFun{\A}{\rad{}}{t} = \frac{\N}{(4\pi \Da
t)^{3/2}}\EXP{\frac{-\radmag{}^2}{4\Da t}}.
\end{equation}

Eq. (\ref{APR12_22}) is the form that is typically used in molecular
communications to describe the local concentration at the receiver;
the receiver is assumed to be a point observer, as in
\cite{RefWorks:498, RefWorks:546}, or
the concentration throughout the receiver volume is assumed to be uniform and
equal to that expected in the center, as in \cite{RefWorks:512}. Eq.
(\ref{APR12_22}) is the baseline against which we will evaluate our proposed
system design.

It has been
noted that Fick's second law violates the theory of special relativity, since
there is no bound on how far a single particle can travel within a given time. A
finite propagation speed can be added as a correction, as in
\cite{RefWorks:435}, but we assume that Fick's second law is sufficiently
accurate without this correction.

\subsection{Reaction-Diffusion}

We now include active enzymes in our analysis.
If we write $\CxFun{\X}{\rad{}}{t} = \Cx{\X}$, $\X
\in \{\A,\En,\EA\}$, for compactness, then the
general reaction-diffusion equation is
\cite[Eq. 8.12.1]{RefWorks:602}
\begin{equation}
\label{JUN12_33}
\pbypx{\Cx{\X}}{t} = \Dx{\X}\nabla^2\Cx{\X} + f\left(\Cx{\X},\rad{},t\right),
\end{equation}
where $f\left(\cdot\right)$ is the reaction term. Using the principles
of chemical kinetics (see \cite[Ch. 9]{RefWorks:585}), we write the complete
partial differential equations for the species in our system as
\begin{align}
\label{AUG12_39}
\pbypx{\Cx{\A}}{t} = &\; \Da\nabla^2\Cx{\A} -\kth{1}\Cx{\A}\Cx{\En} +
\kth{-1}\Cx{\EA}, \\
\label{AUG12_40}
\pbypx{\Cx{\En}}{t} = &\; \De\nabla^2\Cx{\En} -\kth{1}\Cx{\A}\Cx{\En} +
\kth{-1}\Cx{\EA} + \kth{2}\Cx{\EA},\\
\label{AUG12_41}
\pbypx{\Cx{\EA}}{t} = &\; \Da\nabla^2\Cx{\EA} +\kth{1}\Cx{\A}\Cx{\En} -
\kth{-1}\Cx{\EA} - \kth{2}\Cx{\EA}.
\end{align}

This system of equations is highly coupled due to the reaction terms and has no
closed-form analytical solution under our boundary conditions. We seek such a
solution, so we must make some simplifying assumptions. We first note that the
total concentration of enzyme, both free and bound to $\A$, over the entire
system is always constant $\Cx{\Etot} = \Ne/\Ve$.
A common next step for
the Michaelis-Menten mechanism in
(\ref{k1_mechanism})-(\ref{k2_mechanism}) is to assume that the amount of $\EA$
is constant, i.e., $\pbypx{\Cx{\EA}}{t} = 0$, in order to derive an expression
for $\Cx{\EA}$; see \cite[Ch. 10]{RefWorks:585} and its use when
considering enzymes at the receiver in \cite{RefWorks:625}. We will use a
slightly different assumption to directly derive a lower bound
expression. We assume that both $\Cx{\En}$ and $\Cx{\EA}$ are not time-varying,
i.e., $\Cx{\En}$ and $\Cx{\EA}$ are both constants.
It is then straightforward to show that, in our system, (\ref{AUG12_39}) has
solution
\begin{equation}
\label{JUN12_47_proof}
\Cx{\A} \approx \frac{\N}{(4\pi \Da
t)^{3/2}}\EXP{-\kth{1}\Cx{\En}t - \frac{\radmag{}^2}{4\Da t}} +
\kth{-1}\Cx{\EA}t,
\end{equation}
and we ignore (\ref{AUG12_40}) and (\ref{AUG12_41}).
Next, we assume that the amount of $\EA$ at any time is small, such
that $\kth{-1}\Cx{\EA} \to 0$. If
$\Cx{\EA}$ is small, then we can approximate $\Cx{\En}$ with its upper bound
$\Cx{\Etot}$. All concentrations and rate constants must be non-negative, so we
can write the bound
\begin{equation}
\label{JUN12_47}
\Cx{\A} \ge \frac{\N}{(4\pi \Da
t)^{3/2}}\EXP{-\kth{1}\Cx{\Etot}t - \frac{\radmag{}^2}{4\Da t}},
\end{equation}
which is intuitively a lower bound because the actual degradation due to enzymes
can be no more than if all enzymes were always unbound. In other words,
(\ref{JUN12_47}) describes the point concentration of $\A$ molecules as
$\kth{2} \to\infty$ and $\kth{-1}\to 0$. A convenient property of this lower
bound is that, while it loses accuracy as $\EA$ is initially created
($\Cx{\En} < \Cx{\Etot}$),
it eventually improves with time as all $\A$ molecules
are degraded and none remain to bind with the enzyme ($\Cx{\A}, \Cx{\EA} \to 0$,
$\Cx{\En} \to \Cx{\Etot}$, as $t \to \infty$). We also note
that, had we started with the $\pbypx{\Cx{\EA}}{t} = 0$ assumption,
then we would have arrived at a similar expression to (\ref{JUN12_47}), where
$\kth{1}$ is replaced with $\kth{1}\kth{2}/\left(\kth{-1} + \kth{2}\right)$.

Eq. (\ref{JUN12_47}) can be
directly compared with (\ref{APR12_22}). The presence of enzyme results in an
additional decaying exponential term. This decaying exponential is what will
eliminate the ``tail'' that is observed under diffusion alone. It can
be shown that adding enzymes will always lead to a faster degradation of
$\Cx{\A}$ from its maximum value for a given $\rad{}$ than when not adding
enzymes. Furthermore, the maximum value is achieved sooner when enzymes are
present, but this value is smaller. These statements are apparent from the
results in Section~\ref{sec_results}, and future work will prove these
statements analytically.

We have already established that the receiver is able to count the number of
free $\A$ molecules that are within the receiver volume. Eqs. (\ref{JUN12_47})
and (\ref{APR12_22}) give us the expected point concentrations with and without
active enzymes, respectively. We can readily convert these concentrations to the
expected number of observed $\A$ molecules, $\Nobsavgt =
\CxFun{\A}{\rad{0}}{t}\Vobs$, if we assume that the concentration throughout
the receiver is uniform, as in \cite{RefWorks:512}.

\section{Simulation Framework}
\label{sec_sim}

In the previous section, we derived (\ref{JUN12_47}) as a lower bound on the
local concentration when enzymes are present throughout the propagation
environment. We now require
an appropriate simulation framework to evaluate the accuracy of
(\ref{JUN12_47}). This framework will be used to perform stochastic simulations
of the system of equations described by (\ref{AUG12_39})-(\ref{AUG12_41}).

\subsection{Choice of Framework}

Commonly used stochastic reaction-diffusion simulation platforms can be
placed into one of two categories. The first are subvolume-based methods, where
the reaction environment is divided into one (if diffusion is ignored) or many
well-stirred subvolumes. By well-stirred, it is meant that molecules in a
specific subvolume are uniformly distributed throughout that subvolume, and
that the velocities of those molecules follow the Boltzmann distribution; see
\cite{RefWorks:621}. In other words, every subvolume should have more
nonreactive molecular collisions than reactive collisions.
Stochastic subvolume-based methods are based on the stochastic
simulation algorithm, which generates random numbers to determine the time and
type of the next reaction in the system; see \cite{RefWorks:616}. We note that
these methods, though subvolume-based, still consider discrete species
populations.
However, the precise locations of individual molecules are not maintained, and
diffusion is modeled as transitions of molecules between adjacent subvolumes;
see \cite{RefWorks:612}.

The second category of simulation platforms use particle-based methods,
where the precise locations of all individual molecules are known. Every
free molecule diffuses independently along each dimension. These methods require
a constant global time step $\Delta t$ and there is a separation in the
simulation of reaction and diffusion; see \cite{RefWorks:623}. First, all free
molecules are independently diffused along each dimension by generating normal random
variables with variance $2\Dx{\X} \Delta t$. Next, potential reactions are
evaluated to see whether they would have occurred during $\Delta t$. For
bimolecular reactions, a binding radius $\radbind$ is defined as how close
the centers of two reactant molecules need to be at the end of $\Delta t$ in
order to assume that the two molecules collided and bound during $\Delta t$.
For unimolecular reactions, a random number is generated using the rate
constant to declare whether the reaction occurred during $\Delta t$.

Particle-based methods tend to be less computationally efficient, but they do
not have to meet the well-stirred requirement. Our system has an impulse
of molecules being released into an environment with highly reactive enzymes
(we will discuss specific rate constants in Section~\ref{sec_results}, but for
now we note that we are generally interested in large
$\kth{1}$). A general criterion for subvolume size is that the typical
diffusion time for each species should be much less than the typical reaction
time; see \cite{RefWorks:613}. We cannot guarantee the satisfaction of this
criterion for subvolume sizes that make physical sense (i.e., significantly
larger than the size of individual molecules), so we adopt a particle-based
method.

\subsection{Simulating Reactions}

Our bimolecular reaction (\ref{k1_mechanism}) (the binding of $\En$ and
$\A$ to form $\EA$) is reversible, so we must be careful in our choice of
binding radius $\radbind$, time step $\Delta t$, and what we assume
when $\EA$ reverts back to $\En$ and $\A$ molecules. A relevant metric is the
root mean square step length, $\stepl$, between $\En$ and $\A$ molecules, given
as \cite[Eq. 23]{RefWorks:623}
\begin{equation}
\label{AUG12_25}
\stepl = \sqrt{2\left(\Da + \De\right)\Delta t}.
\end{equation}

If reaction (\ref{kminus1_mechanism}) occurs, then the root mean square
separation of the product molecules $\A$ and $\En$ along each dimension is
$\stepl$.
Unless $\stepl \gg \radbind$, then these two reactants will likely
undergo reaction (\ref{k1_mechanism}) in the next time step.
Generally, we need to define an unbinding radius specifying
the initial separation of the $\A$ and $\En$ molecules when reaction
(\ref{kminus1_mechanism}) occurs.
However, in the long time step limit, we can define $\radbind$ as \cite[Eq.
27]{RefWorks:623}
\begin{equation}
\label{AUG12_26}
\radbind = \left(\frac{3\kth{1}\Delta t}{4\pi}\right)^\frac{1}{3},
\end{equation}
and this is valid only when $\stepl \gg \radbind$. Thus, if $\stepl$ is much
greater than $\radbind$ found by (\ref{AUG12_26}), which we can impose by
our selection of $\kth{1}$ and $\Delta t$, then we do not need to
implement an unbinding radius. Also, the evaluation of $\radbind$ is much more
involved when we are not in the long time step limit and requires generating a
lookup table; see \cite{RefWorks:623} for further details. We will select
parameters so that $\stepl \gg \radbind$ is satisfied, even if $\stepl$
becomes comparable with the size of the receiver, so we simply use
(\ref{AUG12_26}).

We have a few additional comments on simulating reaction
(\ref{k1_mechanism}).
It does not require the generation of any random values, besides those that are
used to diffuse the individual molecules. However, we must check the position of
every unbound $\A$ molecule with that of every unbound $\En$ molecule to see
whether they are closer than $\radbind$. For computational efficiency, we create
subvolumes so that we only need to check the positions of enzymes in the current
and adjacent subvolumes of the current free $\A$ molecule. If we find a pair
close enough, then we move both of them to the midpoint of the line between
their centers and re-label them as a single $\EA$ molecule.

Our two unimolecular reactions have the same reactant, $\EA$, so we must
consider both of them when calculating the probability of either
reaction occuring.
For (\ref{kminus1_mechanism}), we have \cite[Eq. 14]{RefWorks:623}
\begin{equation}
\label{AUG12_23}
\Pr\{\textnormal{Reaction (\ref{AUG12_40})}\} =
\frac{\kth{-1}}{\kth{-1}+\kth{2}}
\left[1 - \EXP{-\Delta t \left(\kth{-1}+\kth{2}\right)}\right],
\end{equation}
where $\Pr\{\cdot\}$ denotes probability and (\ref{k2_mechanism}) has
an analogous expression by switching $\kth{-1}$ and $\kth{2}$. A single
random number uniformly distributed between 0 and 1 can then be used to
determine whether a given $\EA$ molecule reacts. If it does, then we place the
products at the same coordinates.

\subsection{Simulating the Sender and Receiver}

When the sender releases an impulse of $\N$ $\A$ molecules, we enforce an
initial separation of $2\Ri{\A}$ between adjacent $\A$ molecules, placing them
in a spherical shape centered at the origin. At the receiver, we make
observations at integer multiples of time step $\Delta t$. When an observation
is made, all free $\A$ molecules whose centres are within $\Vobs$ are counted.

\subsection{Selecting Component Parameters}

We now discuss practical parameter values for the underlying
reaction-diffusion system. Specific enzymatic reactions, such as the breakdown
of acetylcholine by acetylcholinesterase, are represented by
specific molecules and are characterized by specific
reaction rate constants.
Most enzymes are proteins and are usually on the order of less than 10\,nm in
diameter; see \cite[Ch. 4]{RefWorks:588}. From
(\ref{JUN12_60}), smaller molecules diffuse faster, so we are most likely to
select small molecules as information molecules. Many common small
organic molecules, such as glucose, amino acids, and nucleotides, are about 1\,nm
in diameter. In the limit, single covalent bonds between two atoms are about
0.15\,nm long; see \cite[Ch. 2]{RefWorks:588}.

Higher rate constants correlate to faster reactions. Bimolecular rate constants
can be no greater than the collision frequency between the two reactants, i.e.,
every collision results in a reaction. The largest possible value of $\kth{1}$
is on the order of $1.66\times 10^{-19}\,\molecule^{-1}\metre^3\second^{-1}$;
see \cite[Ch. 10]{RefWorks:585} where the limiting rate is listed as on
the order of $10^8\,$L/mol/s. $\kth{2}$ usually varies between 1 and
$10^5\,\second^{-1}$, with values as high as $10^7\,\second^{-1}$. In theory, we
are not entirely limited to pre-existing enzyme-substrate pairs; protein and
ribozyme engineering techniques can be used to modify and optimize the enzyme
reaction rate, specificity, or thermal stability, or modify enzyme function in
the presence of solvents.

\section{Results and Discussion}
\label{sec_results}

We are now prepared to present results comparing the observed number of $\A$
molecules at a receiver with and without the presence of enzymes in the
propagation environment. We assume that the
environment has a viscosity of
$10^{-3}\,\textnormal{kg}\,\metre^{-1}\second^{-1}$ and temperature of
$25\,^{\circ}\mathrm{C}$. The sender emits $\N = 10^4$ $\A$ molecules,
each having radius $\Ri{\A} = 0.5$\,nm, in a single impulse. $\Ve$ is defined as
a cube with side length 1\,$\mu$m and centered at the origin, so its size is on
the order of a bacterial cell.
$\Ne = 2\times
10^5$ $\En$ molecules having radius $\Ri{\En} = 2.5$\,nm are uniformly
distributed throughout $\Ve$. For simplicity, we assume that $\Ri{\EA}
= \Ri{\A} + \Ri{\En} = 3$\,nm. In consideration of the limiting values of
reaction rate constants, we choose $\kth{1} =
10^{-19}\,\molecule^{-1}\metre^3\second^{-1}$, $\kth{-1} = 10^4\,\second^{-1}$,
and $\kth{2} = 10^6\,\second^{-1}$. We also set $\Delta t =
0.5\,\mu\second$, resulting in $\stepl = 22.9$\,nm and $\radbind = 2.28$\,nm, so
that $\stepl \gg \radbind$ is satisfied.

\begin{figure}[!tb]
\centering
\includegraphics[width=\linewidth]{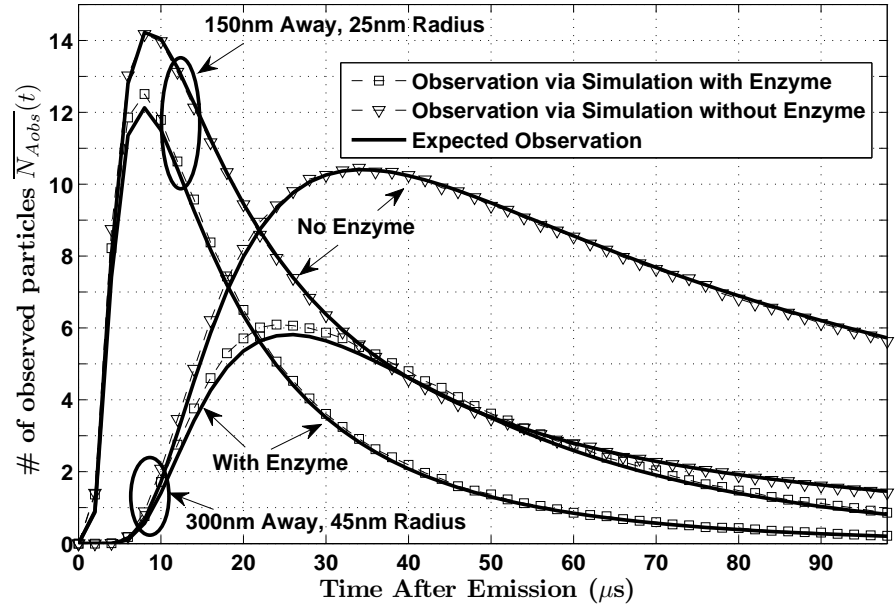}
\caption{Number of particles counted by receivers with radii $\robs =
\{25,45\}$\,nm that are placed $\radmag{0} =
\{150,300\}$\,nm from the sender, respectively.
The source releases $10^4$ molecules at $t
= 0$. Simulation and analytical results are shown both with and without
active enzymes.}
\label{AUG12_27_18}
\end{figure}

We compare the number of molecules observed at a receiver due to a
single emission from the sender.
In Fig.~\ref{AUG12_27_18}, we consider two receivers with radii $\robs =
\{25,45\}$\,nm and their centers placed at a distance of $\radmag{0} =
\{150,300\}$\,nm from the sender, respectively. The expected number of molecules
is calculated using either (\ref{JUN12_47}) or (\ref{APR12_22}) for enzymes
present and absent, respectively. The observed number of $\A$ molecules via
simulation is averaged over at least 15000 independent emissions by the
sender at $t = 0$.

Let us first consider the receiver placed 150\,nm from the sender. The
maximum number of molecules is received about $8\,\mu\second$ after emission.
The maximum value is less than 15\,\% higher in the absence of active enzymes;
over 14 molecules are expected and observed on average via simulation without
enzymes, compared to 12 molecules expected and 12.5 molecules observed with
active enzymes. The decay from the maximum value is slower in the absence of
active enzymes; $60\,\mu\second$ after emission, 3 molecules are expected and
observed without enzymes while 1 molecule is expected and observed with
enzymes, a threefold difference. We see
that, as previously noted, the expected number of observed $\A$ molecules when
active enzymes are present is a lower bound on the average number of $\A$
molecules observed in simulation, and this is a relatively tight bound.

The simulation and analytical results for the receiver placed 300\,nm
from the sender follow the same general trends as those for the closer
receiver, but with a few noteworthy differences. Obviously, the time elapsed
before receiving the maximum number of molecules is greater and the maximum
value is less than for the closer receiver, even though the receiver is larger
(the receiver being larger accounts for how it is possible for this receiver to
observe more molecules than the closer receiver after $23\,\mu\second$).
However, the change in the number of molecules received is much greater in the
presence of active enzymes; the peak number of molecules is observed relatively
sooner (about $25\,\mu\second$ instead of about $35\,\mu\second$ after
emission), but the maximum number of molecules is less than 60\,\% of that
expected without enzymes (about 6 molecules instead of 10.5 molecules).
Intuitively, being further from the sender gives more time for the $\En$
molecules to bind to and then degrade the $\A$ molecules.

Both receivers in Fig.~\ref{AUG12_27_18} show that adding enzymes decreases the
``tail'' of diffusion while still providing a peak to be detected at the
receiver.
It is clear that the sender could emit impulses more often with less risk of
\ISI. For example, if the criterion for designing the bit interval $\T$ was the
time at which the expected number of particles is some fraction of the maximum
expected number, then this time should be shorter in the presence of
active enzymes. Alternatively, sender-receiver pairs could be placed closer
together with less risk of co-channel interference. We leave formal proofs of
these statements for future work, but they are intuitive given the results in
Fig.~\ref{AUG12_27_18}.

\begin{figure}[!tb]
\centering
\includegraphics[width=\linewidth]{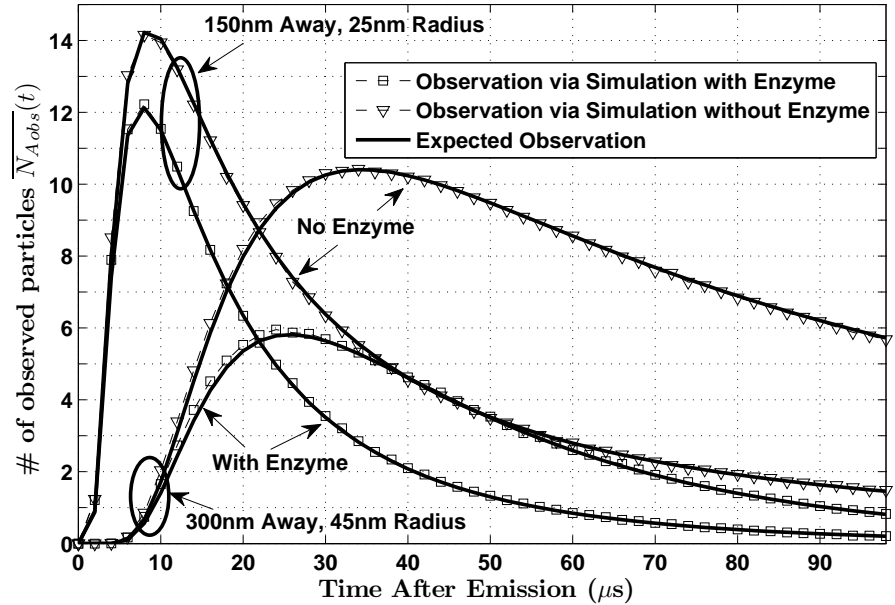}
\caption{Number of particles counted by receivers with radii $\robs =
\{25,45\}$\,nm that are placed $\radmag{0} =
\{150,300\}$\,nm from the sender, respectively.
The source releases $10^4$ molecules from a point at $t
= 0$. Simulation and analytical results are shown both with and without
active enzymes, where $\kth{2} = \infty$ and $\kth{-1} = 0$. The
resultant difference between this figure and Fig.~\ref{AUG12_27_18} is
that here the curves generated via simulation are tighter to the curves
generated by the analytical expressions.}
\label{AUG12_27_20}
\end{figure}

Finally, we consider in Fig.~\ref{AUG12_27_20} the limiting case that
we used to derive the bound (\ref{JUN12_47}), i.e., set $\kth{2} = \infty$,
$\kth{-1} = 0$, and co-locate all $\A$ molecules at the origin when emitting. In
this case, an $\En$ molecule binding to an $\A$ molecule immediately degrades
the $\A$ molecule while releasing the $\En$ molecule, so all enzymes are always
available to react.
We otherwise maintain the same parameters that we used for
Fig.~\ref{AUG12_27_18}. We see that the average number of particles observed via
simulation with active enzymes agrees very well with that expected from
(\ref{JUN12_47}), and that the average number of particles observed via
simulation without active enzymes matches the value expected from
(\ref{APR12_22}), even though we are still assuming uniform $\Cx{\A}$
throughout the receiver volume.
This confirms that the looseness of the lower bound (\ref{JUN12_47}) in
Fig.~\ref{AUG12_27_18} comes from both the finite emission volume and the
creation of $\EA$ molecules. The slight looseness of the lower bound
in Fig.~\ref{AUG12_27_20} for the receiver 300\,nm away and when enzymes are
present is likely due to having a finite $\Ve$; some $\A$ molecules are able to
diffuse beyond $\Ve$, where they cannot be degraded, and then enter the
receiver volume after returning to $\Ve$. This effect is negligible at the
receiver 150\,nm away.

\section{Conclusions and Future Work}
\label{sec_concl}

In this paper, we introduced the concept of using enzymes in the propagation
environment to improve the performance of a diffusive molecular communication
system.
Enzymes that break down information molecules are able to reduce the
time that a sender must wait before being able to send additional information
molecules. There is potential to increase the data rate and to decrease
the probability of error. This gain in performance comes with no additional
complexity required at either the sender or receiver.

The emphasis in this paper was the description of the underlying
reaction-diffusion model and the selection of an appropriate simulation
framework, thereby providing a foundation for performance analysis. On-going
work includes the derivation of the bit error rate for this binary-coded
communication network when multiple emissions are made by the sender given a bit
interval $\T$ and the reception scheme at the receiver. Furthermore, we
are currently evaluating the analytical accuracy of the assumption that the
concentration observed at the receiver is uniform. We must also consider
the ability to choose reaction rate constants based on specific enzymes, as well
as the enzyme concentration. In addition, dimensional
analysis is useful to arbitrarily scale our system,
compare different parameter sets, and derive the looseness
of our receiver bound in terms of a dimensionless parameter. We also note that
we could forego the use of enzymes altogether and use $\A$ molecules with a
faster natural degradation rate, as in \cite{RefWorks:469}, but without using
the number of counted molecules as the amount of information received. This case
would allow simpler and accurate analysis though we would have to be concerned
with maintaining a stockpile of these molecules at the sender without them
degrading before emission. Other relevant problems of interest include
interference from nearby sender/receiver pairs and the potential mobility of
the sender and receiver.

\bibliography{nano_ref}

\begin{thebibliography}{10}

\bibitem{RefWorks:588}
Alberts, B., Bray, D., Hopkin, K., Johnson, A., Lewis, J., Raff, M., Roberts,
  K., Walter, P.:
\newblock Essential Cell Biology. 3rd edn.
\newblock Garland Science (2010)

\bibitem{RefWorks:540}
Akyildiz, I.F., Brunetti, F., Blazquez, C.:
\newblock Nanonetworks: A new communication paradigm.
\newblock Computer Networks \textbf{52}(12) (May 2008)  2260--2279

\bibitem{RefWorks:608}
Nakano, T., Moore, M.J., Wei, F., Vasilakos, A.V., Shuai, J.:
\newblock Molecular communication and networking: Opportunities and challenges.
\newblock IEEE Trans. Nanobiosci. \textbf{11}(2) (Jun. 2012)  135--148

\bibitem{RefWorks:587}
Nelson, P.:
\newblock Biological Physics: Energy, Information, Life. Updated 1st edn.
\newblock W. H. Freeman and Company (2008)

\bibitem{RefWorks:485}
Hiyama, S., Moritani, Y.:
\newblock Molecular communication: Harnessing biochemical materials to engineer
  biomimetic communication systems.
\newblock Nano Commun. Net. \textbf{1}(1) (Mar. 2010)  20--30

\bibitem{RefWorks:512}
Atakan, B., Akan, O.B.:
\newblock Deterministic capacity of information flow in molecular nanonetworks.
\newblock Nano Commun. Net. \textbf{1}(1) (Mar. 2010)  31--42

\bibitem{RefWorks:546}
Mahfuz, M.U., Makrakis, D., Mouftah, H.T.:
\newblock Characterization of intersymbol interference in concentration-encoded
  unicast molecular communication.
\newblock In: Proc. 2011 IEEE CCECE. (May 2011)  164--168

\bibitem{RefWorks:548}
Einolghozati, A., Sardari, M., Beirami, A., Fekri, F.:
\newblock Capacity of discrete molecular diffusion channels.
\newblock In: Proc. 2011 IEEE ISIT. (Aug. 2011)  723--727

\bibitem{RefWorks:557}
Chou, C.T.:
\newblock Molecular circuits for decoding frequency coded signals in
  nano-communication networks.
\newblock Nano. Comm. Networks \textbf{3}(1) (Mar. 2012)  46--56

\bibitem{RefWorks:625}
Nakano, T., Okaie, Y., Vasilakos, A.V.:
\newblock Throughput and efficiency of molecular communication between
  nanomachines.
\newblock In: Proc. 2012 IEEE WCNC. (Apr. 2012)  704--708

\bibitem{RefWorks:513}
Miorandi, D.:
\newblock A stochastic model for molecular communications.
\newblock Nano Commun. Net. \textbf{2}(4) (Dec. 2011)  205--212

\bibitem{RefWorks:469}
Moore, M.J., Suda, T., Oiwa, K.:
\newblock Molecular communication: Modeling noise effects on information rate.
\newblock IEEE Trans. Nanobiosci. \textbf{8}(2) (Jun. 2009)  169--180

\bibitem{RefWorks:581}
Naka, T., Shiba, K., Sakamoto, N.:
\newblock A two-dimensional compartment model for the reaction-diffusion system
  of acetylcholine in the synaptic cleft at the neuromuscular junction.
\newblock Biosystems \textbf{41}(1) (Jan. 1997)  17--27

\bibitem{RefWorks:583}
Cheng, Y., Suen, J.K., Radić, Z., Bond, S.D., Holst, M.J., McCammon, J.A.:
\newblock Continuum simulations of acetylcholine diffusion with
  reaction-determined boundaries in neuromuscular junction models.
\newblock Biophys. Chem. \textbf{127}(3) (May 2007)  129--139

\bibitem{RefWorks:585}
Chang, R.:
\newblock Physical Chemistry for the Biosciences.
\newblock University Science Books (2005)

\bibitem{RefWorks:621}
Gillespie, D.T.:
\newblock A rigorous derivation of the chemical master equation.
\newblock Physica A \textbf{188}(1–3) (Sep. 1992)  404--425

\bibitem{RefWorks:498}
Pierobon, M., Akyildiz, I.F.:
\newblock Information capacity of diffusion-based molecular communication in
  nanonetworks.
\newblock In: Proc. 2011 IEEE INFOCOM 2011. (Apr. 2011)  506--510

\bibitem{RefWorks:435}
Pierobon, M., Akyildiz, I.F.:
\newblock A physical end-to-end model for molecular communication in
  nanonetworks.
\newblock IEEE J. Sel. Areas Commun. \textbf{28}(4) (May 2010)  602--611

\bibitem{RefWorks:602}
Debnath, L.:
\newblock Nonlinear Partial Differential Equations for Scientists and
  Engineers. 2nd edn.
\newblock Birkhaeuser (2005)

\bibitem{RefWorks:616}
Gillespie, D.T.:
\newblock Stochastic simulation of chemical kinetics.
\newblock Annu. Rev. Phys. Chem. \textbf{58}(1) (May 2007)  35--55

\bibitem{RefWorks:612}
Iyengar, K.A., Harris, L.A., Clancy, P.:
\newblock Accurate implementation of leaping in space: the spatial
  partitioned-leaping algorithm.
\newblock J. Chem. Phys. \textbf{132}(9) (Mar. 2010)  094101

\bibitem{RefWorks:623}
Andrews, S.S., Bray, D.:
\newblock Stochastic simulation of chemical reactions with spatial resolution
  and single molecule detail.
\newblock Physical Biology \textbf{1}(3) (Aug. 2004)  137

\bibitem{RefWorks:613}
Bernstein, D.:
\newblock Simulating mesoscopic reaction-diffusion systems using the
  {G}illespie algorithm.
\newblock Phys. Rev. E \textbf{71}(4) (Apr. 2005)  041103

\end{thebibliography}
\bibliographystyle{splncs}

\end{document}